\documentclass{aa}  
\usepackage{graphicx}
\usepackage{float}
\usepackage{txfonts}
\usepackage{xcolor}

\newcommand{\feh}{[\element{Fe}/\element{H}]}
\newcommand{\fpl}{\bar{n}_\mathrm{p}}
\newcommand{\Npl}{n_\mathrm{p}}
\newcommand{\Ntotpl}{N_\mathrm{p}}
\newcommand{\Neff}{N^\mathrm{eff}_\mathrm{s}}
\newcommand{\Ns}{N_\mathrm{s}}

\newcommand{\NtotplHost}{N_\mathrm{h}}
\newcommand{\FplH}{F_\mathrm{h}}

\begin{document}

   \title{Occurrence rates of small planets from HARPS  \thanks{Based on observations collected at the European Organization for Astronomical Research in the Southern Hemisphere under ESO programs (see \emph{acknowledgements} for a full list of used programs).}}

   \subtitle{Focus on the Galactic context}

   \author{D.~Bashi\inst{1} 
            \and 
          S.~Zucker\inst{1} 
            \and 
          V.~Adibekyan\inst{2,3} 
          \and 
          N.~C.~Santos\inst{2,3} 
          \and
          L.~Tal-Or\inst{4}
          \and
          T.~Trifonov\inst{5} 
          \and  
          T.~Mazeh\inst{6}}

   \institute{Porter School of the Environment and Earth Sciences, Raymond and Beverly Sackler Faculty of Exact Sciences, Tel Aviv University, Tel Aviv, 6997801, Israel. 
    \email{dolevbas@mail.tau.ac.il} 
    \and
    Instituto de Astrofísica e Ciências do Espaço, Universidade do Porto, CAUP, Rua das Estrelas, PT4150-762 Porto, Portugal 
    \and 
    Departamento de Física e Astronomia, Faculdade de Ciências, Universidade do Porto, Rua do Campo Alegre, 4169-007 Porto, Portugal
    \and 
    Department of Physics, Ariel University, Ariel 40700, Israel
    \and 
    Max-Planck-Institut für Astronomie, Königstuhl 17, D-69117 Heidelberg, Germany 
    \and 
    School of Physics and Astronomy, Raymond and Beverly Sackler Faculty of Exact Sciences, Tel Aviv University, Tel Aviv, 6997801, Israel }
             
\titlerunning{Planet occurrence rates from HARPS}
\authorrunning{Bashi et al.}

   \date{Received ...; accepted ...}

 
  \abstract
   {The stars in the Milky Way thin and thick disks can be distinguished by several properties such as metallicity and kinematics. It is not clear whether the two populations also differ in the properties of planets orbiting the stars. In order to study this, a careful analysis of both the chemical composition and mass detection limits is required for a sufficiently large sample. Currently, this information is still limited only to large radial-velocity (RV) programs. Based on the recently published archival database of the High Accuracy Radial velocity Planet Searcher (HARPS) spectrograph, we present a first analysis of low-mass (small) planet occurrence rates in a sample of thin- and thick-disk stars.}
   {We aim to assess the effects of stellar properties on planet occurrence rates and to obtain first estimates of planet occurrence rates in the thin and thick disks of the Galaxy. As a baseline for comparison, we also aim to provide an updated value for the small close-in planet occurrence rate and compare it to results of previous RV and transit (\textit{Kepler}) works.}
   {We used archival HARPS RV datasets to calculate detection limits of a sample of stars that were previously analysed for their elemental abundances. For stars with known planets we first subtracted the Keplerian orbit. We then used this information to calculate planet occurrence rates according to a simplified Bayesian model in different regimes of stellar and planet properties.}
   {Our results suggest that metal-poor stars and more massive stars host fewer low-mass close-in planets. We find the occurrence rates of these planets in the thin and thick disks to be comparable. In the iron-poor regimes, we find these occurrence rates to be significantly larger at the high-$\alpha$ region (thick-disk stars) as compared with the low-$\alpha$ region (thin-disk stars). In general, we find the average number of close-in small planets ($2$--$100$\,days, $1$--$20$\,$M_{\oplus}$) per star (FGK-dwarfs) to be: $\fpl = 0.36 \pm 0.05,$ while the fraction of stars with planets is $\FplH = 0.23 \substack{+0.04 \\ -0.03}$. Qualitatively, our results agree well with previous estimates based on RV and \textit{Kepler} surveys. }
   {This work provides a first estimate of the close-in small planet occurrence rates in the solar neighbourhood of the thin and thick disks of the Galaxy. It is unclear whether there are other stellar properties related to the Galactic context that affect small-planet occurrence rates, or if it is only the combined effects of stellar metal content and mass. A future larger sample of stars and planets is needed to address those questions.}


   \maketitle
%

%

\section{Introduction}

The study of the population of small exoplanets 
and its statistical relation with the population of their host stars is substantive in order to understand planet formation and evolution. Results from radial-velocity (RV) spectrographs such as the High Resolution Echelle Spectrometer \citep[HIRES;][]{Vogetal1994} and the High Accuracy Radial Velocity Planet Searcher \citep[HARPS;][]{Pepe03, Mayor03}, with a precision on the order of $1\,\mathrm{m\,s}^{-1}$, as well as results from the \textit{Kepler} mission, confirmed the existence of hundreds of small planets and multi-planet systems. Unlike giant planets around FGK-dwarfs that are believed to form preferentially around metal-rich stars \citep[e.g.,][]{Santos04, FischerValenti, Johnson10}, first estimates suggest that small planets can be found orbiting stars with a wide range of metallicities \citep{Sousa08,Buchhave}. Subsequent works analysing small-planet occurrence rate (henceforth SPOR) tried to estimate the fraction of Sun-like stars that harbour Earth-like planets \citep[e.g.,][]{Youdin11, Burke15, Hsu19, Zhu}, investigated the effects of \feh\ on small planets \citep[e.g.,][]{Petigura, Zhu, Bashi19}, and revealed a radius gap between super-Earth and sub-Neptune planets \citep{Fulton}. Both RV and transit surveys have shown that close-in compact multi-planet systems are usually composed of planets smaller than Neptune and that Jovian planets (especially the hot Jupiters) are uncommon \citep{Winn}.

In the Galactic context, it is not clear how the stellar birthplace affects the likelihood of forming and maintaining planets. The stars in the solar neighbourhood are commonly grouped into three main populations: the thin and thick disks, and the halo \citep{Gilmore83}. Considering disk stars, the properties of the local thick-disk population have been characterised by many previous spectroscopic studies. Thick-disk stars are commonly assumed to be older, kinematically hotter, and more iron-poor and $\alpha$-enhanced than the thin-disk stars \citep[e.g.,][]{Gilmore89, Reddy06, Adibekyana, Bensby14, Buder18, Bashi19}. In an attempt to relate planetary structure and composition to stellar elemental abundances, \cite{Santos17} showed that disks around stars affiliated to different Galactic populations can form rocky and water planets with significantly different iron-to-silicate mass fractions. Consequently, these latter authors argued that their results may have a significant impact on our understanding of the occurrence rate of planets in the Galaxy.

Several works \citep{Haywood,Gonzalez,Adibekyanb} suggested that while planets can be found in iron-poor regimes, their occurrence is linked to the presence of other metals (especially $\alpha$ elements). However, these studies did not take into account mass detection limits in their estimates of planet occurrence rate.

The recent information from \textit{Gaia} DR2 \citep{Gaia18}, combined with ground-based spectroscopic surveys, has already improved our understanding of the differences among the solar neighbourhood stellar populations in the Galactic context \citep[e.g.,][]{Hayes18, Buder18}. Unfortunately, planet search RV programs are still very limited to the solar neighbourhood; see for example the HARPS GTO samples \citep{Mayor03, LoCurto, Santos11}, which we use in this work. With the combined available elemental abundance estimates of many stars in the solar neighbourhood \citep{Sousa08, Adibekyana, DelgadoMena17} and the recent public release of a large fraction of the HARPS RVs by \citet{Trifonov20}, we are now able to make updated estimates of planet occurrence in a well-defined and large RV sample. As the HARPS GTO stars were selected from a volume-limited stellar sample observed by CORALIE \citep[typical precision $5$--$10$ $\mathrm{{m\,s}^{-1}}$;][]{Udry2000}, our analysis is particularly valid for the estimation of SPOR for $M_\mathrm{p}\sin i$ lower than $30\,M_{\oplus}$, that is, below the CORALIE detection limit \citep{ Sousa08}.

In this work, we aim to examine the SPOR in the Galactic context, and specifically its affiliation with the thin and thick disks. We restricted our analysis to disk stars and ignored the rare population of halo stars in the solar neighbourhood. 

The following section describes the way we compiled our samples of stars and planets. Section~\ref{Occurrence} presents the method we used to calculate planet occurrence rates. We present our results in Section~\ref{Results} and discuss our findings in Section~\ref{Discussion}.

\section{The samples} \label{The Sample}

\subsection{Stellar sample}

Our starting point was a sample of $1111$ FGK stars observed in the context of the HARPS GTO programs \citep{Mayor03, LoCurto, Santos11}. Previous works used high signal-to-noise-ratio HARPS spectra of those stars to estimate chemical abundances of $25$ elements \citep{Sousa08, Adibekyana, DelgadoMena17}, including \element{Fe} and some $\alpha$ elements such as \element{Mg}, \element{Si} and \element{Ti} as well as the stellar masses. As we focused on FGK dwarfs, we excluded stars with $\log g$ smaller than $4.0$. We then cross-matched this sample with the recently published HARPS RV dataset \citep{Trifonov20}\footnote{https://www2.mpia-hd.mpg.de/homes/trifonov/HARPS\_RVBank.html}, which provided RV estimates and other data products for all available HARPS targets. For our mass detection-limit calculations we chose to keep only stars with more than $12$ observations\footnote{A somewhat arbitrary number of observations, nevertheless enough for possible detection of close-in planets.}. Thus, our final sample includes $693$ stars.

\begin{figure}
   \centering
   \includegraphics[width=10cm]{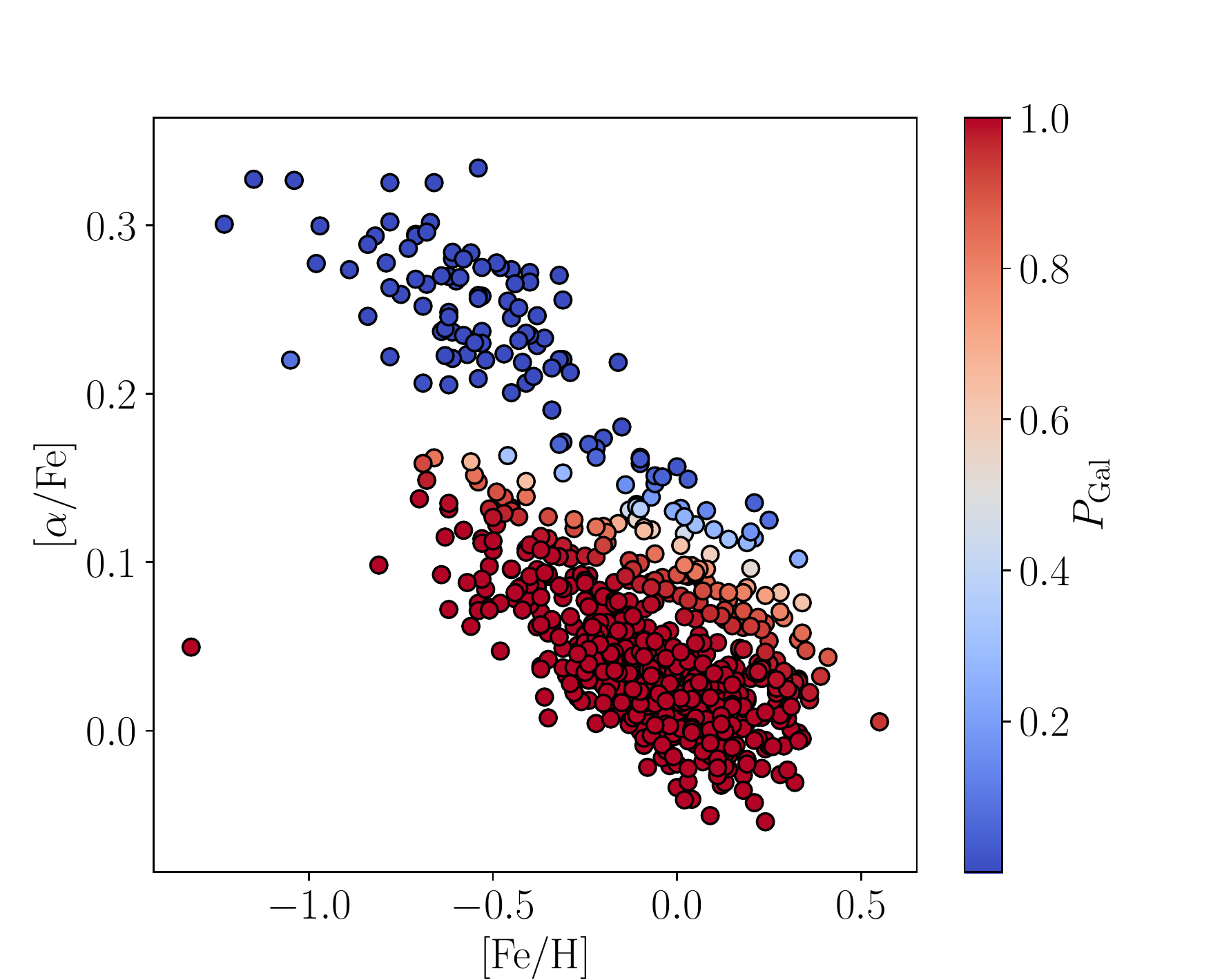}
      \caption{Scatter plot of the stellar sample on the [$\alpha$/\element{Fe}]-\feh\ plane. The colour coding represents the probability $P_{\mathrm{Gal}}$ of affiliation to each Galactic cluster according to the Gaussian Mixture Model where red ($=1$) represents thin disk and blue ($=0$) thick disk.}
         \label{GalacticContextScatter}
   \end{figure}

Figure~\ref{GalacticContextScatter} shows our stellar sample on the [\element{Fe}/\element{H}]-[$\alpha$/\element{Fe}] plane. Following \cite{Blancato}, we used a Gaussian Mixture Model (GMM) and fixed the number of clusters to two in order to distinguish between the thin- and thick-disk stars. The figure presents the results of the GMM analysis by colour coding the probability $P_{\mathrm{Gal}}$ that a point belongs to one of the two clusters. Cutting at a probability of $0.5$, the cluster marked with red corresponds to the thin disk, while the blue cluster to the thick disk. This procedure reveals that $573$ stars in our sample are affiliated with the thin disk, and $120$ stars with the thick disk. The shapes of the two clusters we find broadly agree with the results of previous works on stellar populations \citep{Reddy06, Adibekyanb, Bensby14, Blancato}. We note that in our GMM analysis, we assumed a two-population model, of thin- and thick-disk stars. It is still not clear whether the h$\alpha$mr population \citep{Adibekyan11} is a separate one or is simply the metal-rich tail of the thick disk. As the technique we used to separate the populations is not the same as in the work of \cite{Adibekyanb}, the stellar Galactic context affiliation we find is not completely identical to that of these later authors, although it is based on a subgroup of stars listed in that work.

\subsection{Planet sample}

In order to build our planet sample, we reanalysed the available HARPS data in a consistent way that would account for the mass detection limits. As most of the publicly available HARPS RVs have already been carefully scrutinised for planetary signals, our aim is not to perform a manual search and confirmation of new planetary signals. Instead, we decided to rely on publicly available exoplanet catalogues. Following \cite{Bashi18}, the main exoplanet catalogue we used was the NASA Exoplanet Data Archive \citep{Akeetal2013}. We also included planets listed in the Extrasolar Planets Encyclopaedia \citep{Schetal2011} and the sample of \cite{Mayor11}. For each planet-host star we used the corresponding published planetary orbits as starting points for our analysis. We first repeated the Keplerian orbit fitting, because the HARPS-RVBANK data somewhat differ from the original velocities that were used for the published solution.

As correction of RV signals originating from stellar activity, rotation, or sampling is often a subjective process that might be prone to errors, we decided not to correct for those effects and not to apply such manipulations to the RV dataset \citep{Faria16}. Consequently, similarly to \cite{Faria16}, our following detection-limit estimates can be considered conservative.

We define the mass detection limit for a given star as the upper limit on the mass ($M_\mathrm{p}\sin i$) that could not have been detected at a given period given the RV uncertainties and the epochs. We computed the detection limits using the Local Power Analysis (LPA) method \citep{Meunier}, assuming circular orbits\footnote{As previous RV works have suggested, the detectability is only slightly affected by this assumption for eccentricities below $0.5$ \citep{Endetal2002,Mayor11}.}, on a grid of $150$ logarithmically-spaced orbital periods in the range $2$--$3000$\,days. The LPA method compares the maximum power $P_\mathrm{inj}$ of the RV signal induced by injecting a mock planetary signal of a given mass and period (with the same temporal sampling as the actual data) to the maximum power $P_\mathrm{act}$ of the actual RV signal within a local period range of the periodogram. In our case, we applied the LPA method with a similar set of parameters as suggested in \cite{Borgetal2017}. The injected planetary signal was deemed to be above the detection limit if, for $12$ consecutive different phase realisations, $P_\mathrm{inj}$ was always above $P_\mathrm{act}$. For each trial orbital period, we used a binary search on a grid of $M_\mathrm{p}\sin i$ where the upper limit was $13\,M_\mathrm{J}$, and the finest grid step was $0.3\,M_{\oplus}$. 

We performed this procedure for each star in our stellar sample to derive a planet detection probability: $p_{\mathrm{S}}$ (an example is presented in Fig. \ref{DetectionProbFig}). We defined the planet detection probability of a star as the ratio between the area above its mass detection limit curve and the total area of a rectangle of logarithmic planetary period and $M_\mathrm{p}\sin i$. We note that in doing so, we implicitly assumed that the prior probability distributions of the period and $M_\mathrm{p}\sin i$ are log-uniform.. 

When a target was found to have a companion, based on our cross-match with the exoplanet catalogues, we first subtracted the RV planetary signal (by fitting a Keplerian orbit) and only afterwards calculated the detection limits. Our final planet sample included only planets that were found to be above the mass detection limit. Overall, we found almost all publicly available confirmed planets to be above the detection limit curve. The only exceptions were: (i) planets found only after applying intense and dedicated statistical treatment to the data, such as for example binning, moving average, or Gaussian process analysis, to overcome stellar activity and correlated noise: HD\,10700\,g,\,h \citep{Feng17b}, HD\,40307\,c \citep{Tuomi13}, HD\,177565\,b \citep{Feng17a}, HD\,215152\,b,\,c \citep{Delisle18}; (ii)  HD\,125612\,c \citep{Ment18} which was detected with data obtained mainly with other instruments, that is, not HARPS; and (iii) transiting planets detected recently by TESS \citep{Ricker} that could not be detected with the available archival HARPS data sets: GJ\,143\,c \citep{Dragomir19}, HD\,15337\,b,\,c \citep{Gandolfi19}, and HD\,23472\,b,\,c \citep{Trifonov19}.

\begin{figure}
\centering
 \includegraphics[width=9.5cm]{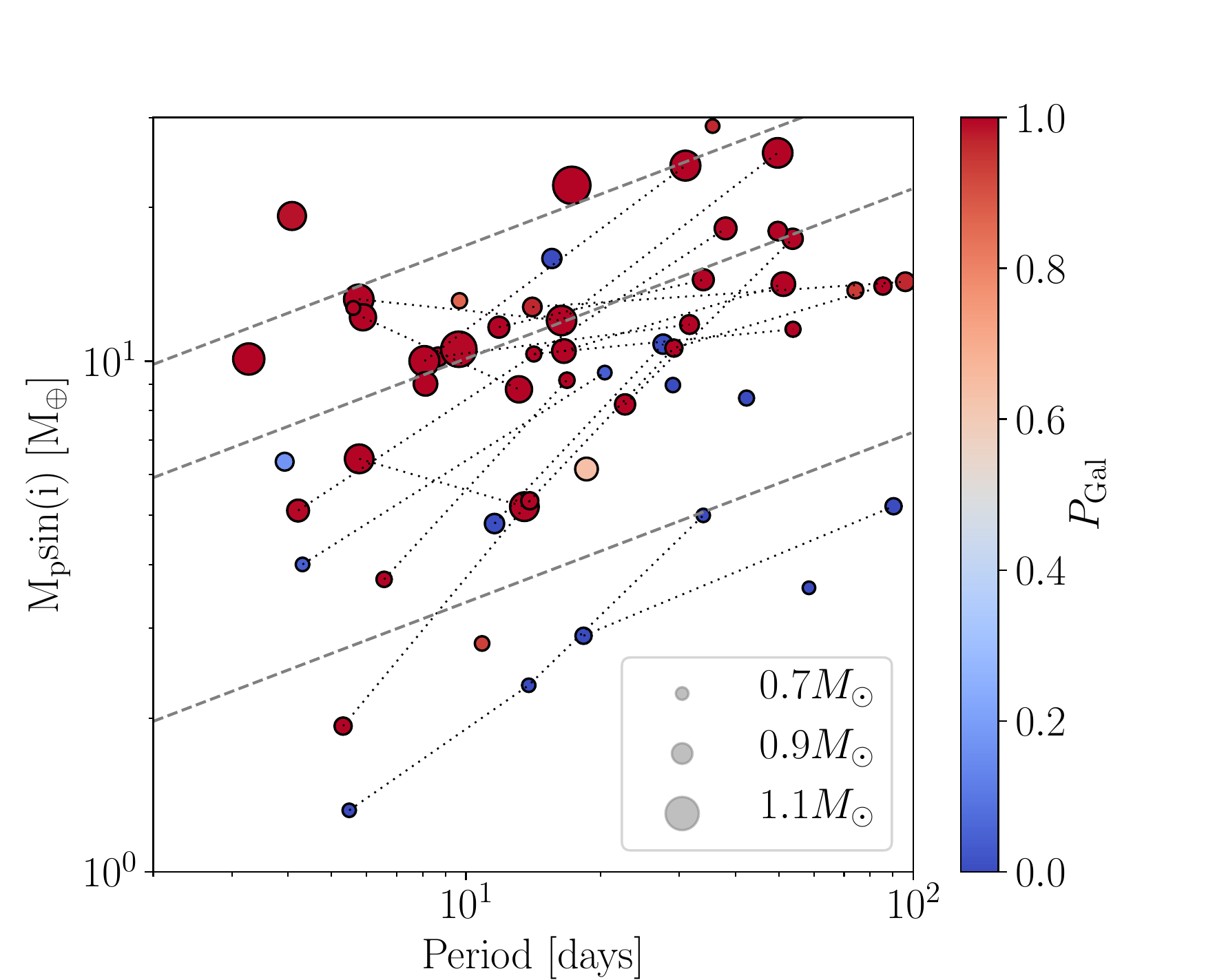}
  \caption{Planet sample on the period--mass plane, with thin- and thick-disk affiliation colour-coded in the same way as in Fig.~\ref{GalacticContextScatter}. The circle sizes represent the masses of the host stars. Planets in the same system are connected by dotted lines. The dashed lines represent the RV detection threshold for circular planetary signals corresponding to semi-amplitudes of $1$, $3,$ and $5$ $\mathrm{m~ s^{-1}}$ (from bottom to top), for a $1$~$M_{\mathrm{\odot}}$ star.}
     \label{Scat_M-Per}
\end{figure}

In total, we identified $55$ close-in ($P~ \mathrm{<}~100$ days) small planets ($M_\mathrm{p}\sin i < 30\,M_{\oplus}$) orbiting $34$ stars (Table.~\ref{PlanetList}, Fig.~\ref{Scat_M-Per}) that were found to be above the detection limit. As suggested, all planets used in this work are based on published works.
\begin{center}
\begin{table*}[ht]
\caption{Planet and host parameters for each of the 55 planets we identified. The table lists planet orbital period and $M_\mathrm{p}\sin i$, host-star parameters mass, [Fe/H], $\mathrm{[\alpha /Fe]}$ \citep{Sousa08, Adibekyana, DelgadoMena17}, the Galactic affiliation probability $P_{\mathrm{Gal}}$ (i.e. probability of affiliation to the thin disk $P_{\mathrm{Gal}} \sim 1 $ or thick disk $P_{\mathrm{Gal}} \sim 0 $), and $P_\mathrm{S}$ planet detection probability.}
\label{PlanetList} 
\centering
\begin{tabular}{lcccccccc} 
\hline\hline
 Planet Name & Period [days] & $M_\mathrm{p}\sin i~[M_{\oplus}]$ & Host Mass $[M_{\odot}]$& [Fe/H] & $\mathrm{[\alpha /Fe]}$ & $P_{\mathrm{Gal}}$ & $P_\mathrm{S}$ \\
\hline

HD 1461 b&$5.77$&$6.44$&$1.04$&$0.19$&$0.02$&$0.99$&$0.65$\\
HD 1461 c&$13.51$&$5.19$&$1.04$&$0.19$&$0.02$&$0.99$&$0.65$\\
HD 4308 b&$15.56$&$15.89$&$0.87$&$-0.34$&$0.22$&$3.92\times 10^{-4}$&$0.38$\\
HD 10180 c&$5.76$&$13.22$&$1.05$&$0.08$&$0.03$&$0.99$&$0.67$\\
HD 10180 d&$16.36$&$12.01$&$1.05$&$0.08$&$0.03$&$0.99$&$0.67$\\
HD 10180 e&$49.75$&$25.59$&$1.05$&$0.08$&$0.03$&$0.99$&$0.67$\\
HD 13808 b&$14.18$&$10.33$&$0.77$&$-0.20$&$0.07$&$0.99$&$0.42$\\
HD 13808 c&$53.83$&$11.55$&$0.77$&$-0.20$&$0.07$&$0.99$&$0.42$\\
HD 16417 b&$17.24$&$22.10$&$1.15$&$0.13$&$0.03$&$0.99$&$0.46$\\
HD 20003 b&$11.85$&$11.66$&$0.91$&$0.04$&$0.02$&$0.99$&$0.51$\\
HD 20003 c&$33.92$&$14.44$&$0.91$&$0.04$&$0.02$&$0.99$&$0.51$\\
HD 20781 b&$5.31$&$1.93$&$0.83$&$-0.11$&$0.06$&$0.99$&$0.85$\\
HD 20781 c&$13.89$&$5.33$&$0.83$&$-0.11$&$0.06$&$0.99$&$0.85$\\
HD 20781 d&$29.16$&$10.61$&$0.83$&$-0.11$&$0.06$&$0.99$&$0.85$\\
HD 20781 e&$85.51$&$14.03$&$0.83$&$-0.11$&$0.06$&$0.99$&$0.85$\\
HD 20794 b&$18.32$&$2.90$&$0.80$&$-0.40$&$0.27$&$6.02\times 10^{-7}$&$0.74$\\
HD 20794 d&$90.31$&$5.20$&$0.80$&$-0.40$&$0.27$&$6.02\times 10^{-7}$&$0.74$\\
HD 21693 b&$22.68$&$8.23$&$0.89$&$0.00$&$0.04$&$0.99$&$0.46$\\
HD 21693 c&$53.74$&$17.37$&$0.89$&$0.00$&$0.04$&$0.99$&$0.46$\\
HD 21749 b&$35.59$&$28.87$&$0.73$&$-0.02$&$0.08$&$0.96$&$0.30$\\
HD 26965 b&$42.38$&$8.47$&$0.77$&$-0.31$&$0.26$&$3.36\times 10^{-6}$&$0.68$\\
HD 31527 b&$16.55$&$10.47$&$0.96$&$-0.17$&$0.05$&$0.99$&$0.81$\\
HD 31527 c&$51.21$&$14.16$&$0.96$&$-0.17$&$0.05$&$0.99$&$0.81$\\
HD 39194 b&$5.48$&$1.32$&$0.72$&$-0.61$&$0.28$&$4.03\times 10^{-7}$&$0.84$\\
HD 39194 c&$13.81$&$2.32$&$0.72$&$-0.61$&$0.28$&$4.03\times 10^{-7}$&$0.84$\\
HD 39194 d&$33.92$&$4.99$&$0.72$&$-0.61$&$0.28$&$4.03\times 10^{-7}$&$0.84$\\
HD 40307 b&$4.31$&$4.10$&$0.73$&$-0.31$&$0.17$&$0.05$&$0.51$\\
HD 40307 d&$20.43$&$9.50$&$0.73$&$-0.31$&$0.17$&$0.05$&$0.51$\\
HD 45184 b&$5.89$&$12.19$&$1.00$&$0.04$&$0.03$&$0.99$&$0.41$\\
HD 45184 c&$13.14$&$8.81$&$1.00$&$0.04$&$0.03$&$0.99$&$0.41$\\
HD 47186 b&$4.08$&$19.25$&$1.03$&$0.23$&$0.04$&$0.98$&$0.22$\\
HD 51608 b&$14.07$&$12.77$&$0.86$&$-0.07$&$0.09$&$0.96$&$0.70$\\
HD 51608 c&$95.94$&$14.31$&$0.86$&$-0.07$&$0.09$&$0.96$&$0.70$\\
HD 69830 b&$8.67$&$10.20$&$0.86$&$-0.06$&$0.06$&$0.99$&$0.55$\\
HD 69830 c&$31.60$&$11.80$&$0.86$&$-0.06$&$0.06$&$0.99$&$0.55$\\
HD 85512 b&$58.43$&$3.60$&$0.70$&$-0.32$&$0.27$&$6.32\times 10^{-7}$&$0.93$\\
BD-08 2823 b&$5.60$&$12.71$&$0.74$&$-0.06$&$0.05$&$0.99$&$0.19$\\
HD 90156 b&$49.77$&$17.98$&$0.86$&$-0.24$&$0.07$&$0.99$&$0.51$\\
HD 96700 b&$8.12$&$9.03$&$0.96$&$-0.18$&$0.05$&$0.99$&$0.63$\\
HD 109271 b&$8.07$&$10.01$&$1.06$&$0.10$&$0.00$&$0.99$&$0.36$\\
HD 109271 c&$30.94$&$24.15$&$1.06$&$0.10$&$0.00$&$0.99$&$0.36$\\
HD 115617 b&$4.21$&$5.11$&$0.92$&$-0.02$&$0.06$&$0.99$&$0.75$\\
HD 115617 c&$38.05$&$18.23$&$0.92$&$-0.02$&$0.06$&$0.99$&$0.75$\\
HD 125595 b&$9.67$&$13.13$&$0.78$&$0.08$&$0.09$&$0.86$&$0.35$\\
HD 134060 b&$3.27$&$10.1$&$1.07$&$0.14$&$0.00$&$0.99$&$0.62$\\
HD 136352 b&$11.58$&$4.81$&$0.87$&$-0.34$&$0.19$&$6.94\times 10^{-3}$&$0.8$\\
HD 136352 c&$27.58$&$10.8$&$0.87$&$-0.34$&$0.19$&$6.94\times 10^{-3}$&$0.8$\\
HD 154088 b&$18.60$&$6.15$&$0.94$&$0.28$&$0.08$&$0.64$&$0.64$\\
HD 160691 d&$9.64$&$10.55$&$1.13$&$0.30$&$0.01$&$0.99$&$0.17$\\
HD 175607 b&$29.01$&$8.98$&$0.75$&$-0.62$&$0.25$&$3.02\times 10^{-5}$&$0.66$\\
HD 176986 b&$6.56$&$3.74$&$0.78$&$0.00$&$0.01$&$0.99$&$0.54$\\
HD 176986 c&$16.82$&$9.18$&$0.78$&$0.00$&$0.01$&$0.99$&$0.54$\\
HD 192310 b&$74.21$&$13.76$&$0.80$&$-0.04$&$0.09$&$0.93$&$0.72$\\
HD 215152 d&$10.86$&$2.80$&$0.76$&$-0.10$&$0.10$&$0.92$&$0.66$\\
HD 215497 b&$3.93$&$6.36$&$0.84$&$0.20$&$0.12$&$0.17$&$0.61$\\

\end{tabular}
\end{table*}
\end{center}

\section{Occurrence-rate estimates} \label{Occurrence}
As was recently pointed out by \citet{Zhu}, there is a distinction between two notions of planet occurrence rate: (i) the average number of planets per star, $\fpl$ 
, and (ii) the fraction of stars with planets,  $\FplH$. The ratio between these two numbers gives the average multiplicity: the average number of planets per planetary system (not per star). Studying both occurrence rate and multiplicity may help to disentangle the effects of planet formation and dynamical evolution. Occurrence-rate studies based on \textit{Kepler} \citep{Petigura, Hsu18} usually address $\fpl$, by counting the number of {detected planets} and dividing it by the number of stars in the survey around which such planets could have been detected. Most of the Doppler surveys on the other hand tend to consider the second notion of occurrence rate, by counting the number of {planet-host stars} and dividing it, again, by the number of stars in the survey around which the planets could have been detected. However, there have  also been a few attempts to apply the $\FplH$ approach to the \textit{Kepler} survey; see for example \cite{Zhu}. 

In any case, before estimating the occurrence rate, it is imperative to first correct the samples for incompleteness. To do so, we used a simplified Bayesian model (SBM), as suggested by \cite{Hsu18}.  The details of the model we used are presented in Appendix \ref{AppendixA} and Appendix \ref{AppendixB} where we describe the method used to estimate $\fpl$ and $\FplH$ respectively. Henceforth, the values we report for planet occurrence rates are the median and the $16$\%-$84$\% ($1\sigma$) confidence interval.  

\section{Results} \label{Results}
We analysed the SPOR in the period--mass ($M_\mathrm{p}\sin i$) range of $2$--$100$ days and $1$--$20\,M_{\oplus}$, that is, super-Earths and sub-Neptune planets. We found the fraction of stars with small planets to be: $\FplH = 0.23 \substack{+0.04 \\ -0.03}$, and the average number of planets per star to be: $\bar{n}_{\mathrm{p}} = 0.36 \pm 0.05$. Our results agree with the results of previous RV works \citep{Howard, Mayor11} that estimated the fraction of stars with planets. This can be seen in Table~\ref{RVcompare}, where we compare our results with those of these latter authors in the same range of parameter space. Furthermore, our results also agree with previous \textit{Kepler}-based works that estimated the average number of planets per star \citep[][see Table~\ref{Keplercompare}]{Fulton, Hsu19}. In the following sections, we investigate the effects of stellar properties on the planet occurrence rates, and present estimates of planet occurrence rates in the thin and thick disks. 

\begin{figure*}
 \centering
\includegraphics[width=8cm]{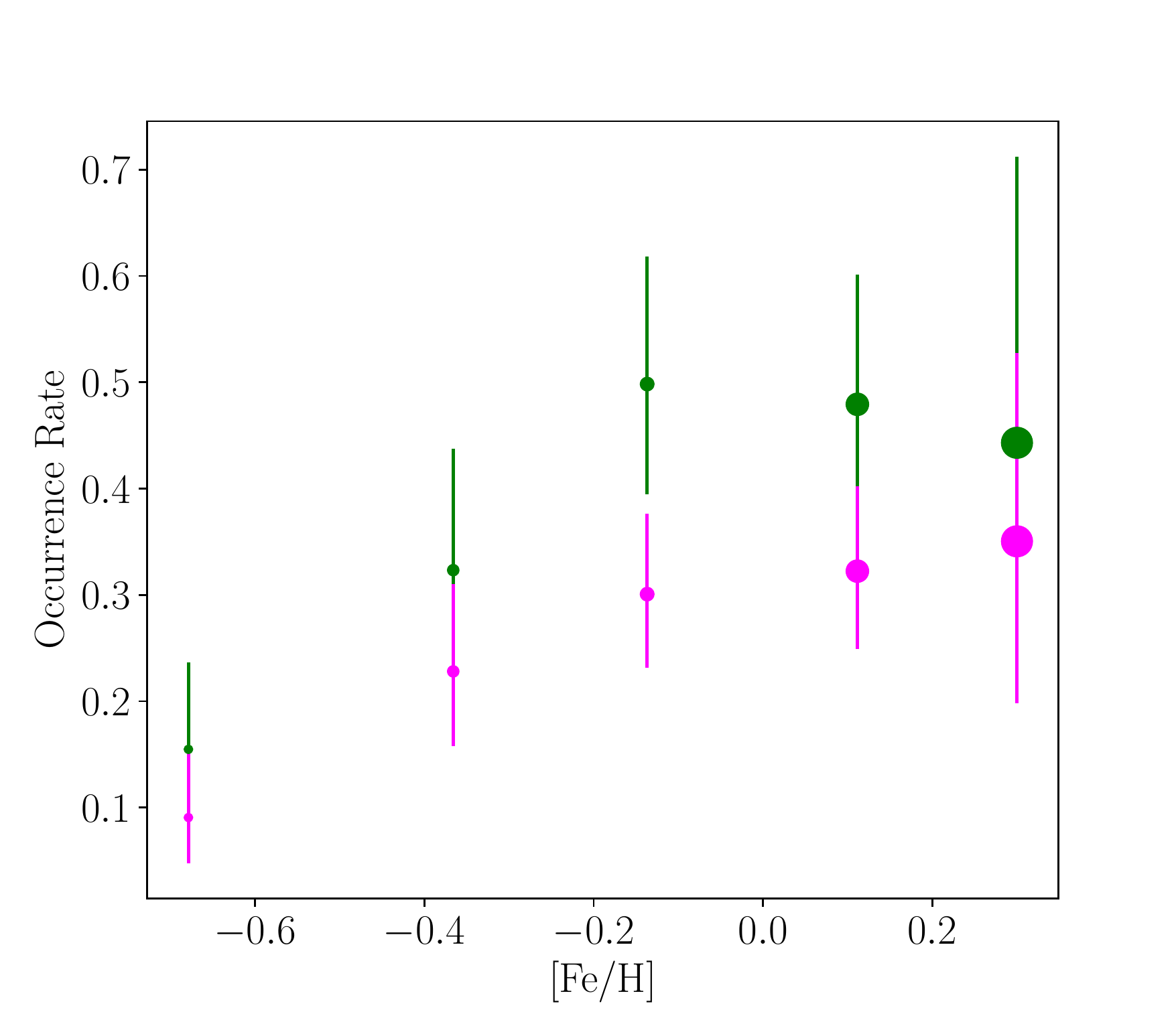}
\includegraphics[width=8cm]{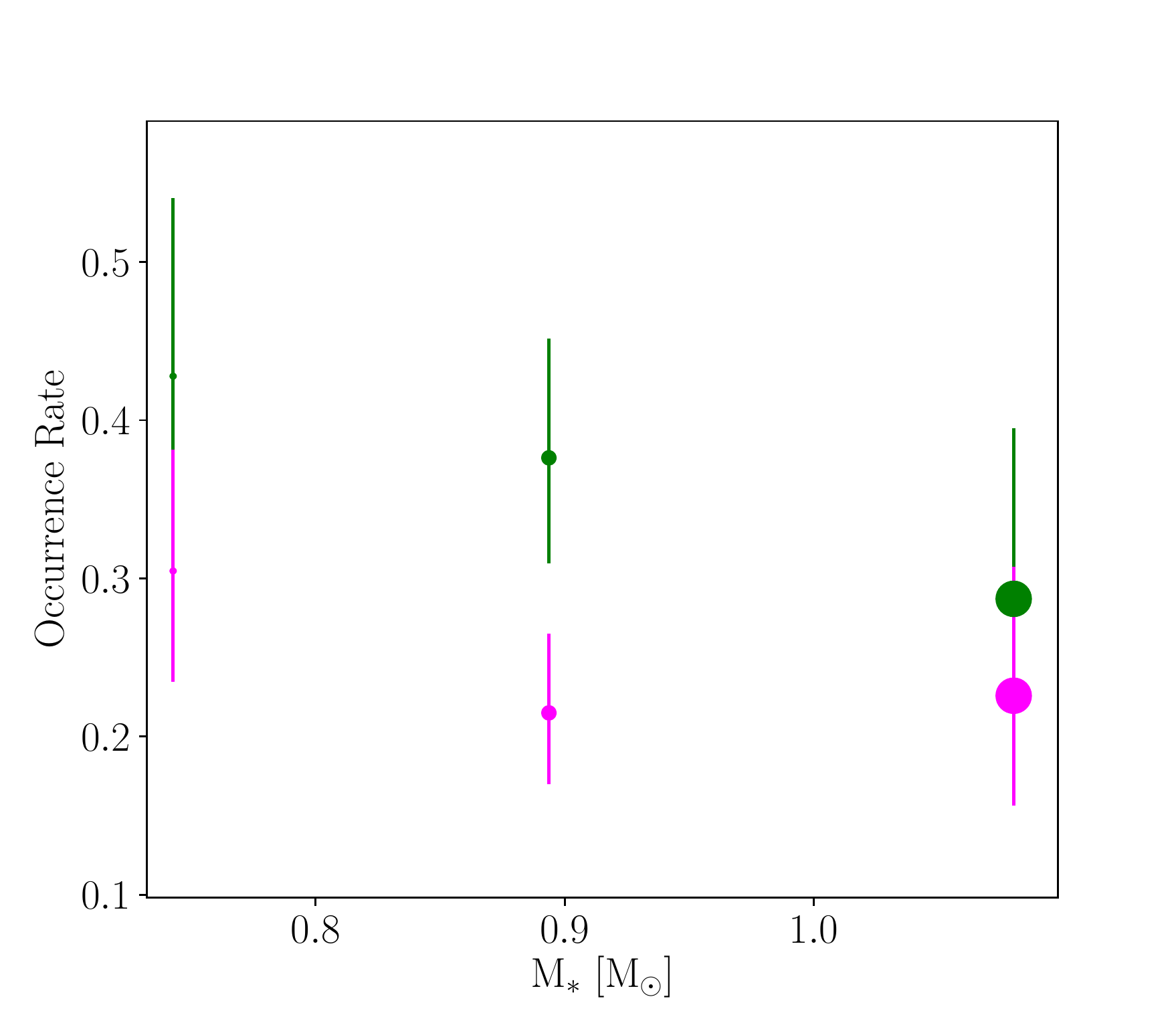}
\caption{Close-in ($P~ \mathrm{<}~100$ days) small planet  ($M_\mathrm{p}\sin i < 30\,M_{\oplus}$) occurrence rates: average number of planets per star ($\fpl$) in green, fraction of stars with planets ($\FplH$) in magenta as a function of \feh\ (left panel) and stellar mass (right panel). The circle sizes represent the mean stellar mass of the relevant bin.} 
\label{FreqPlot}
\end{figure*}

\begin{table}[H]
\caption{\label{RVcompare} Comparison of the fraction of stars with planets ($\FplH$) for periods shorter than $50$ days from various works.}
\centering
\begin{tabular}{lccc}
\hline\hline
  & 3-10 $\mathrm{M_{\oplus}}$&10-30 $\mathrm{M_{\oplus}}$\\
\hline
\cite{Howard}  & $0.12 \pm 0.04$ &$ 0.07 \pm 0.03$ & \\
\cite{Mayor11}  & $0.17 \pm 0.04 $  &$0.11 \pm 0.02$& \\
\hline
This Work\tablefootmark{a}&  $0.13 \pm 0.02$ &$0.10 \pm 0.02$\\
This Work\tablefootmark{b} &$0.11 \pm 0.03$&$0.08 \pm 0.02$&\\
This Work\tablefootmark{c} &$0.10 \pm 0.02$&$0.06 \pm 0.01$&\\
\hline
\end{tabular}
\tablefoot{
\tablefoottext{a}{Based on a HARPS volume-limited sample by \cite{Sousa08}. Occurrence-rate estimate is based on the IDEM method used both in \cite{Howard} and \cite{Mayor11}.}
\tablefoottext{b}{Based on a HARPS volume-limited sample by \cite{Sousa08}. Occurrence-rate estimate is based on the SBM method.}
\tablefoottext{c}{Based on our sample of $693$ stars. Occurrence-rate estimate is based on the SBM method.}
}
\end{table}

\begin{table}[H]
\caption{\label{Keplercompare} Qualitative comparison of recent $\fpl$ estimates for periods between $1$ and $100$ days}
\centering
\begin{tabular}{lccc}
\hline\hline
  & Planetary & $\fpl$ Estimation & $\fpl$\\
  & size range & Method & \\
\hline
\cite{Fulton}&  $2$-$4$ $\mathrm{R_{\oplus}}$  &IDEM& $0.37 \pm 0.02$ \\
\cite{Hsu19}& $2$-$4$ $\mathrm{R_{\oplus}}$  &ABC& $\sim 0.41$\tablefootmark{*}\\
This Work  & $1$-$20$ $\mathrm{M_{\oplus}}$ &SBM& $0.36\pm 0.05$\\
\hline
\end{tabular}
\tablefoot{
\tablefoottext{*}{Value deduced from Fig.~2. of \cite{Hsu19}.}
}
\end{table}

\subsection{Effects of stellar properties}

The left panel of Fig.~\ref{FreqPlot} depicts the effect of stellar metallicity on planet occurrence rates. Below solar iron content, the SPOR seems to be increasing with \feh. For higher \feh,\ nothing conclusive can be inferred because of the large uncertainties. We repeated our analysis for different binning schemes to make sure that the trends we report are not due to a specific arbitrary choice of binning. Assuming our small-planet sample is equivalent to a sample of transiting planets with radii smaller than $4R_{\oplus}$, our results are also consistent with \textit{Kepler} results \citep{Petigura, Zhu}. 

The right panel of Fig.~\ref{FreqPlot} shows the effect of stellar mass. The figure suggests there might be a slight decrease in SPOR with stellar mass. Previous works suggested similar trends based on the dependence of the planet-radius distribution on the stellar spectral type \citep{Mulders}. In a recent work comparing \textit{Kepler}'s planet occurrence rates in different ranges of planet properties, between M- and FGK-dwarfs, \cite{Hsu20} found close agreement in planet occurrence rates when using an equivalent insolation flux, suggesting stellar irradiance plays a significant role in planet evolution processes. However, the apparent trend would be difficult to test with the current small sample. 

\begin{figure}
\centering
\includegraphics[width=8cm]{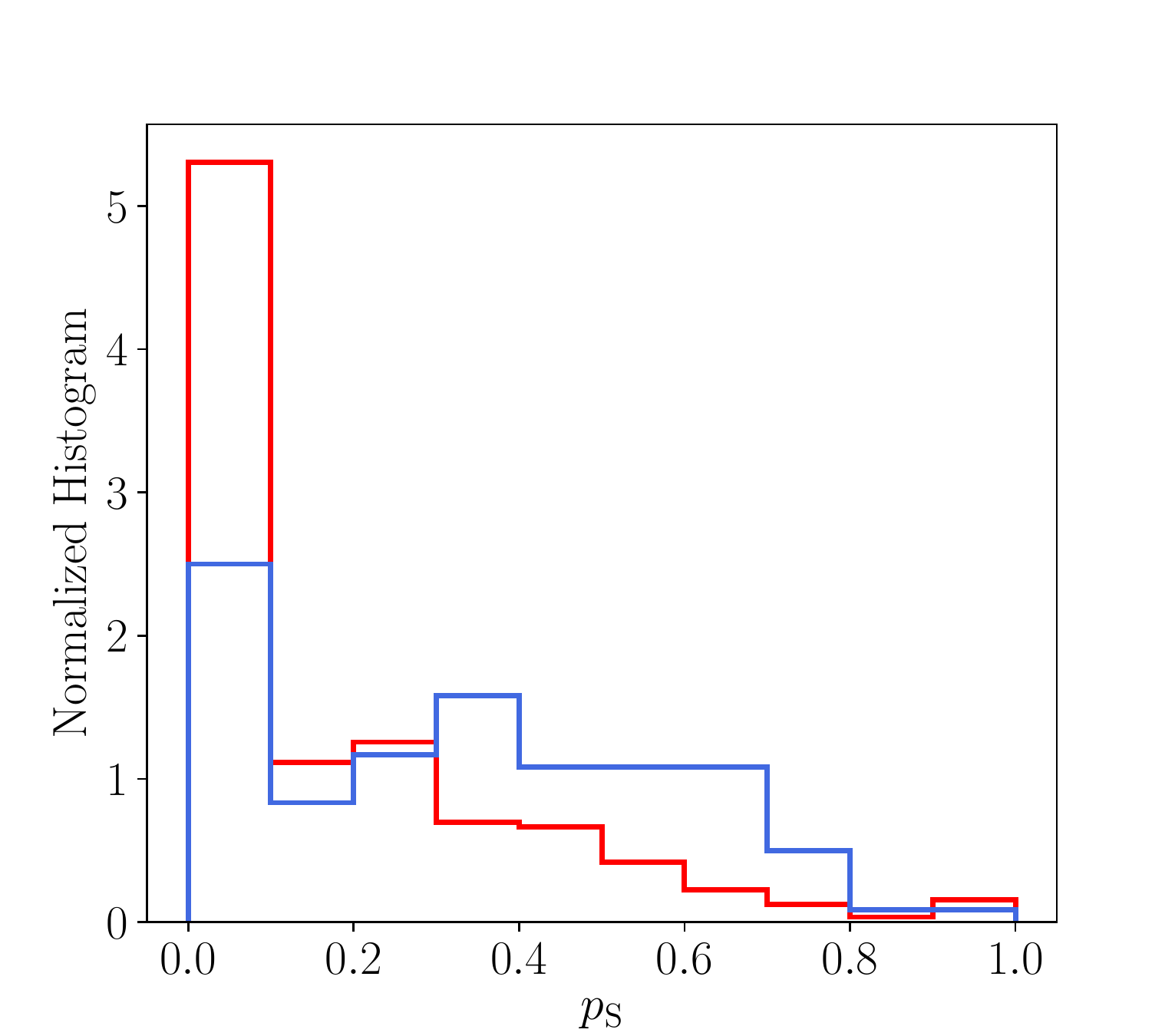}
  \caption{Normalised histogram of planet detectability $p_\mathrm{S}$ for thin-disk (red) and thick-disk (blue) stars.}
     \label{DetectabilityHist}
\end{figure}

\subsection{Thin and thick disks}
In this section, we give first estimates of the close-in SPOR in the thin and thick disks of the Milky Way. We find the occurrence rate in the two samples to be essentially identical: $\bar{n}_{\mathrm{p}} = 0.37 \substack{+0.06 \\ -0.05}$, $F_{\mathrm{p}} = 0.23 \pm 0.04$ in the thin disk; $\bar{n}_{\mathrm{p}} = 0.35 \substack{+0.10 \\ -0.09}$, $F_{\mathrm{p}} = 0.24 \substack{+0.07 \\ -0.06}$ in the thick disk. 

As suggested above, it seems that both stellar metallicity and (possibly) stellar mass affect planet occurrence rates. Most thin-disk stars are rich in iron and more massive\footnote{This might be a selection bias in the sample: (i) metal-poor stars have on average lower masses for a fixed $B-V$ \citep{Santos03}, and (ii) as thick-disk stars are usually older, most massive stars (especially F-type stars) have already evolved away from the main sequence while the younger thin-disk stars have not.} than thick-disk stars. Consequently, putting aside metallicity effects, smaller planets should be somewhat more easily detectable around thick-disk stars as they are less massive compared to their thin-disk counterparts. Figure~\ref{DetectabilityHist} presents a normalised  histogram of planet detectability for thin-disk and thick-disk stars, which suggests that planet detectability among thick-disk stars is higher than that among thin-disk stars. This might introduce some bias that may also explain Fig.~\ref{Scat_M-Per} where most thick-disk stars also host smaller planets than their thin-disk counterparts (however, see \cite{Adibekyan13} and \cite{Sousa19}). 

\begin{figure*}
   \centering
   \includegraphics[width=16cm]{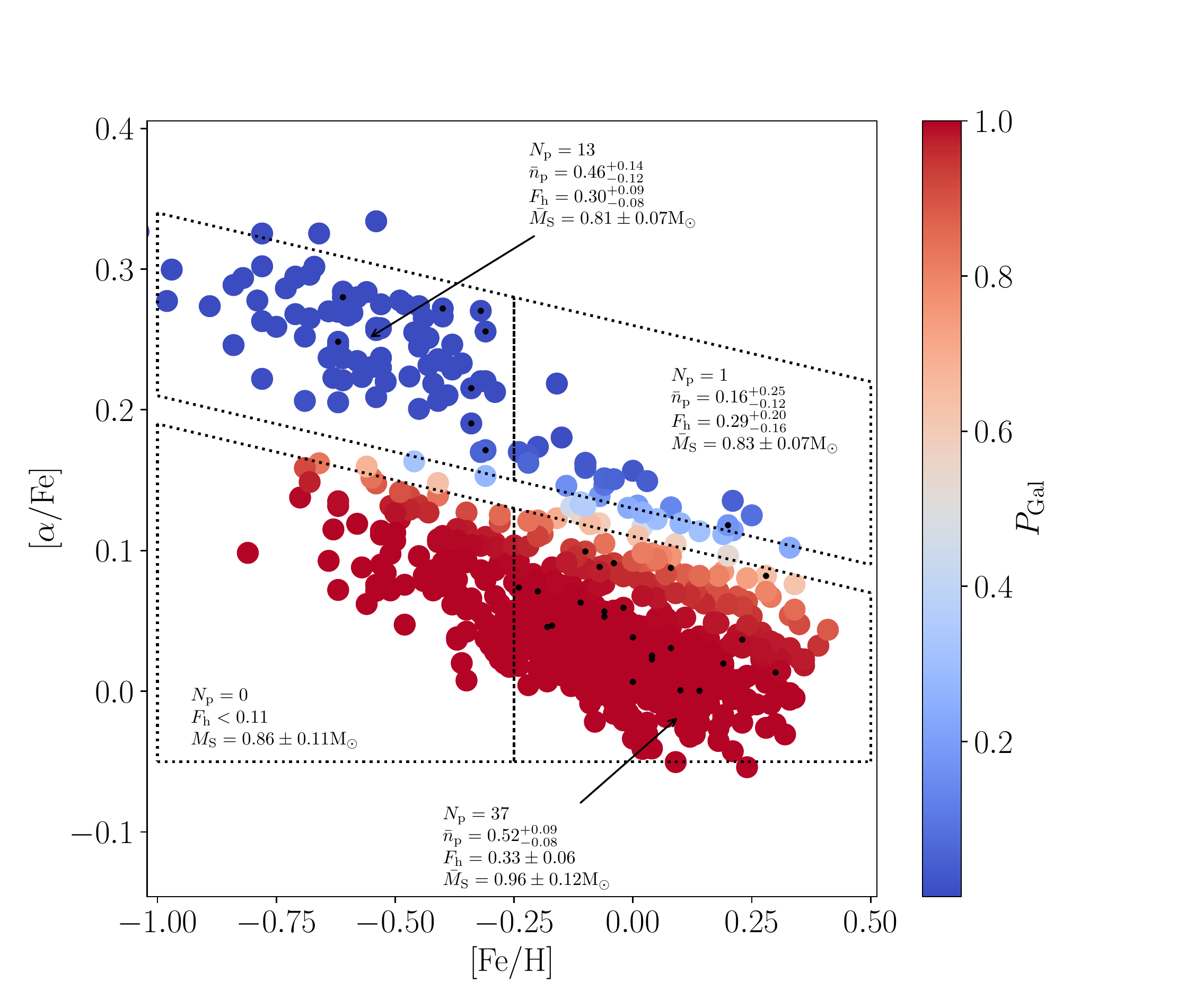}
      \caption{Scatter plot of the stellar sample on the [$\alpha$/\element{Fe}]-\feh\ plane (similar to Fig.1). Red circles represent thin-disk stars while blue circles represent thick-disk stars. Close-in small planet-host stars are marked by black dots. The division into four regions is based on cutting at $\text{\feh}=-0.25$\,dex in addition to the division into thin and thick disks. For each region, we have marked the number and occurrence rates of detected close-in small planets as well as the detectability-weighted means of the stellar mass sample.}
         \label{Occ_GalacticConBins}
   \end{figure*}

In Fig.~\ref{Occ_GalacticConBins} we have divided the \feh-[$\alpha$/\element{Fe}] plane into four regions by cutting at $\text{\feh}=-0.25$ dex \citep [the expected iron-content value where the thin-disk and thick-disk populations start to diverge in $\alpha$ content;][]  {Adibekyana}, in addition to the separation between thin and thick disks we had found with the GMM. We ignored a few stars whose affiliation was ambiguous. In each of the four bins, we estimated  $\fpl$ and $\FplH$ separately. We also calculated the detectability-weighted means of the stellar mass. In light of the division into the four regions, an interesting trend emerges: there are no close-in small planets detected around stars in the low-$\alpha$ region. This precludes a proper estimation of $\fpl$, as is explained in Appendix \ref{AppendixA}. However, adopting the value $0.36$ for $\fpl$, the probability of obtaining zero planets around every star in this bin is $10^{-4}$. According to the posterior distribution of $\FplH$, at a $95\%$ confidence level $\FplH < 0.11$, which is significantly lower than the  value quoted above for the whole sample, namely of $0.23$.  In any case, the absence of any planets around stars in this bin is a highly significant result, supporting claims by \citet{Adibekyana} and \citet{Adibekyanc}. This effect does not seem to be directly related to stellar mass as there is no significant difference between the two regions in this respect. In the iron-rich region, it seems that there is almost no difference in the fraction of stars with planets between the thin- and thick-disk (h$\alpha$mr) samples, yet the current h$\alpha$mr sample is still too small to make any decisive conclusions.

\section{Discussion} \label{Discussion}
This work is based on the recent release of almost all the HARPS RV datasets by \cite{Trifonov20}. We used their HARPS-RVBANK data and revisited previous estimates of the SPOR and its relation to stellar properties and Galactic context. %

Our results provide a first estimate of the SPOR in the solar neighbourhood of the thin and thick disks of the Galaxy. Previous works in this field did not take into consideration the detection limits of each star in performing their planet occurrence-rate calculations. We estimated SPOR in the range $P = 1$--$100$\,days and $M_\mathrm{p}\sin i=1$--$20\,M_{\oplus}$  (i.e. super-Earths and Sub-Neptune planets) by applying a simplified Bayesian model (SBM). 

We used SBM in order to correct the samples for incompleteness and estimate the SPOR, but there are other possible approaches. One popular approach is the Inverse Detection Efficiency Method \citep[IDEM; e.g.][]{Howard12,ForemanMac14}. In this approach, one assigns a weight to each detected planet in order to account for similar planets that had not been detected. With the SBM approach we used, we calculated the detection probability of planets (for a specified range of period and mass) around each individual star. We then divided the number of detected planets or planet-host systems by the effective number of stars in the sample around which planets could have been found. On the one hand, this approach is less sensitive to errors in planet properties. \cite{Hsu18} showed that the occurrence rates computed using SBM agree with the occurrence rates estimated using a full Bayesian model or an approximated Bayesian computation (ABC) model, which they eventually used themselves. On the other hand, it is important to note that our assumption of log-uniform distributions in planet period and $M_\mathrm{p}\sin i$ might eventually result in overestimation of the detection efficiency and thus underestimation of the occurrence rates.

The planet-host fraction we estimate is broadly consistent with the results of the previous works of \cite{Howard} and \cite{Mayor11} which was also based on the HARPS targets. However, it is important to note a few differences that might affect our final results. Firstly, as our sample also includes metal-poor stars of the HARPS-4 subsample \citep{Santos11}, the stars in our sample are on average more metal poor than the stars in the sample of \citet{Mayor11}. Secondly, it seems that  the studies of \cite{Howard} and \cite{Mayor11}  have both overestimated $\FplH$ (the fraction of stars with planets). The probable reason is that when they applied the IDEM method, they used the probability of detecting the `least detectable' planet (hence a higher weight) in a planetary system as the probability of detecting at least one planet. Although less severe for giant planets because of their low multiplicity rate, this overestimate might significantly reduce the estimated fraction of stars with lower-mass planets \citep{Zhu}. 

To put this in perspective, we repeated our analysis in a similar parameter range to that of \cite{Howard} and \cite{Mayor11}. We summarise our results and compare them in Table~\ref{RVcompare} for similar period and mass ranges to those used by \cite{Howard} and \cite{Mayor11}. Reassuringly, when we used the HARPS volume-limited sample of \citet{Sousa08}, the estimated occurrence rates using the IDEM method were compatible with those of \cite{Mayor11}. However, when we used the SBM method, the estimated fraction of stars with planets was lower.

Another interesting comparison is between the SPOR in general, and in a subsample of systems without an outer giant planet. Having found no small candidate planet among $20$ solar-type stars known to host a single long-period giant planet, \cite{Barbato} estimated the fraction of stars with low-mass planets in the presence of an outer giant planet to be $\FplH < 0.10$. Relying on results from \cite{Mayor11}, Barbato et al.,  concluded that the fraction of small planet-host stars with known outer giant planets is much lower than previous estimates of the fraction of small planet-host systems in general. However, \citeauthor{Barbato} compared the two fractions in two different ranges of planet properties.\ While \cite{Mayor11} used the range of periods shorter than $50$~days and masses smaller than $30\,M_{\oplus}$, \citeauthor{Barbato} examined periods shorter than $150$~days and masses in the range $10$--$30\,M_{\oplus}$. These are two different ranges. Indeed, when we repeat our estimates in a range similar to the one \citeauthor{Barbato} assumed, we find $\FplH = 0.07\pm 0.02$, in agreement with their upper limit, suggesting that the frequency of small planets in systems with outer giant planets might be as common as, or perhaps more common than \citep{ZhuWu18, Bryanetal2019} the frequency of small planets in any other systems.

As for the average number of planets per star, we find it to be broadly consistent with recent \textit{Kepler} works (Table~\ref{Keplercompare}), under reasonable assumptions regarding the planetary mass--radius relation \citep[e.g.][]{Bashietal17, Ulmertal2019, Otegietal20}. However, it is important to note that solar-type stars monitored by \textit{Kepler} reside in a different Galactic region (Galactic latitudes $b\sim 5^{\circ}$--$20^{\circ}$). Therefore, their metallicity distribution might be different as compared to a volume-limited sample of solar-type stars. This calls for extra caution when performing such comparisons.

In general, the significant difference between our estimate of the planet-host fraction and the average number of planets per star allows us to estimate the known average number of planets per planetary system, that is, the average multiplicity \citep{Zhu}. Our results suggest that the average multiplicity of close-in small planets  (the ratio  $\fpl/\FplH$) is $ \sim 1.6$, which implies that stars tend to host more than one planet in that mass range. 

Our results on the dependence of planet occurrence rates on stellar \feh\ or mass (Fig. \ref{FreqPlot}) suggest similar trends to those found by previous studies using \textit{Kepler} data \citep{Mulders, Petigura, Zhu}. It seems that there is a metal content threshold under which the probability of forming small and short-period planets is negligible \citep[\text{\feh} < $-0.8$;][]{Mortier16}. This indeed may support current theories of planet formation by solid accretion \citep{Nimmo18}. Nevertheless, we find that in the most metal-rich regimes, the SPOR seems to be lower than for solar metallicity. Previous works suggested this effect might be a consequence of giant outer planets perturbing the formation or survival of possible inner smaller planets \citep{Barbato, Zhu}. We cannot confirm this hypothesis; instead, our results may hint that this effect is directly related to the higher stellar mass in metal-rich stars. An alternative explanation might be related to the stellar irradiance that might evaporate and reduce the planet mass, as was recently suggested by \cite{Hsu20}. In summary, our results confirm that the relations between stellar properties and SPOR seen in the \textit{Kepler} field of view also hold in the solar neighbourhood.

Due to the diverse metallicity and stellar mass distributions, we conclude that although the occurrence rates in the thin and thick disks are compatible, it is difficult at this stage to draw any concrete conclusions about the effect of the Galactic context on SPOR. Focusing on iron-poor stars (the two left segments in Fig.~\ref{Occ_GalacticConBins}), our results suggest a significantly higher SPOR among thick-disk stars as compared to thin-disk stars \cite[see also][]{Adibekyanc}. This result is not evident in the iron-rich bins comparing thin-disk stars with h$\alpha$mr stars. 

As our sample includes both thin- and thick-disk stars with a wide range of iron content and $\alpha$-enhancement values, it is important to distinguish between stars according to the total heavy element content. At a given \feh,\, an $\alpha$-enhanced star will contain a higher fraction of metals. Consequently, one can argue that the higher SPOR found in iron-poor thick-disk stars, as compared to thin-disk stars, might be related to a higher fraction of metals in thick-disk stars as compared to their thin-disk counterparts. To test this, we examined two separate properties instead of metallicity: the iron content (\feh), which is usually used as a proxy to the overall metallicity of a star, and the summed mass fraction of all heavy elements ($Z$) assumed to be needed for forming planets \citep{Santos17}\footnote{We used the approximate relation suggested by \cite{Santos17} to estimate the value of $Z$ in solar-neighbourhood stars using the abundances of \element{Mg}, \element{Si,} and \element{Fe}.}. We calculated the distribution of the detectability-weighted means of  $Z$ and found it to be comparable in the iron-poor thin- ($Z=1.01 \pm 0.09\%$) and thick-disk ($Z=1.00 \pm 0.13\%$) regions. Consequently, we may argue that the significantly higher SPOR among thick-disk stars as compared to thin-disk stars is not caused by a metallicity effect as one might suspect. Alternatively, this could suggest that some elements are particularly relevant for planet formation.

As a final remark, we note that our analysis did not consider the uncertainties in \feh\ and [$\alpha$/\element{Fe}]. However, as those uncertainties are small (less than $0.1$\,dex) for most stars, the results are not expected to be dramatically affected.

Future works should extend this analysis to the \textit{Kepler} survey, as was attempted by \cite{Bashi19} based on analysis of stellar kinematics and \feh. The recent release of LAMOST DR5, and data-driven methods such as for example the Payne model \citep{Xiang} that estimates the elemental content of a large portion of \textit{Kepler} field-of-view stars, will allow further analysis of planet occurrence rates in the Galactic context. Furthermore, the growing sample of planets detected by \textit{TESS} will facilitate similar tests on a large all-sky sample of stars and advance the study of exoplanets in the Galactic context.

\begin{acknowledgements}
We thank the anonymous referee for providing helpful comments that improved the quality of the paper. S.Z. acknowledges the support by the ISRAEL SCIENCE FOUNDATION (grant No. 848/16) and partial support by the Ministry of Science, Technology and Space, Israel. N.C.S acknowledges the support from FCT - Fundação para a Ciência e a Tecnologia through national funds and by FEDER through COMPETE2020 - Programa Operacional Competitividade e Internacionalização by these grants: UID/FIS/04434/2019; UIDB/04434/2020; UIDP/04434/2020; PTDC/FIS-AST/32113/2017 \& POCI-01-0145-FEDER-032113; PTDC/FIS-AST/28953/2017 \& POCI-01-0145-FEDER-028953. V.A. acknowledges the support from FCT through Investigador FCT contract nr. IF/00650/2015/CP1273/CT0001.
The work is based on observations collected
at the European Organization for Astronomical Research in the Southern Hemisphere under ESO programs: 0100.C-0097, 0100.C-0111, 0100.C-0414, 0100.C-0474, 0100.C-0487, 0100.C-0750, 0100.C-0808, 0100.C-0836, 0100.C-
0847, 0100.C-0884, 0100.C-0888, 0100.D-0444, 0100.D-0717, 0101.C-0232, 0101.C-0274, 0101.C-0275, 0101.C-0379, 0101.C-0407, 0101.C-0516, 0101.C- 0829, 0101.D-0717, 0102.C-0338, 0102.D-0717, 0103.C-0548, 0103.D-0717, 060.A-9036, 060.A-9700, 072.C-0096, 072.C-0388, 072.C-0488, 072.C-0513, 072.C-0636, 072.D-0286, 072.D-0419, 072.D-0707, 073.A-0041, 073.C-0733, 073.C-0784, 073.D-0038, 073.D-0136, 073.D-0527, 073.D-0578, 073.D-0590, 074.C-0012, 074.C-0037, 074.C-0102, 074.C-0364, 074.D-0131, 074.D-0380, 075.C-0140, 075.C-0202, 075.C-0234, 075.C-0332, 075.C-0689, 075.C-0710, 075.D-0194, 075.D-0600, 075.D-0614, 075.D-0760, 075.D-0800, 076.C-0010, 076.C-0073, 076.C-0155, 076.C-0279, 076.C-0429, 076.C-0878, 076.D-0103, 076.D-0130, 076.D-0158, 076.D-0207, 077.C-0012, 077.C-0080, 077.C-0101, 077.C-0295, 077.C-0364, 077.C-0530, 077.D-0085, 077.D-0498, 077.D-0633, 077.D-0720, 078.C-0037, 078.C-0044, 078.C-0133, 078.C-0209, 078.C-0233, 078.C-0403, 078.C-0751, 078.C-0833, 078.D-0067, 078.D-0071, 078.D-0245, 078.D-0299, 078.D-0492, 079.C-0046, 079.C-0127, 079.C-0170, 079.C-0329, 079.C-0463, 079.C-0488, 079.C-0657, 079.C-0681, 079.C-0828, 079.C-0927, 079.D-0009, 079.D-0075, 079.D-0118, 079.D-0160, 079.D-0462, 079.D-0466, 080.C-0032, 080.C-0071, 080.C-0664, 080.C-0712, 080.D-0047, 080.D-0086, 080.D-0151, 080.D-0318, 080.D-0347, 080.D-0408, 081.C-0034, 081.C-0119, 081.C-0148, 081.C-0211, 081.C-0388, 081.C-0774, 081.C-0779, 081.C-0802, 081.C-0842, 081.D-0008, 081.D-0065, 081.D-0109, 081.D-0531, 081.D-0610, 081.D-0870, 082.B-0610, 082.C-0040, 082.C-0212, 082.C-0308, 082.C-0312, 082.C-0315, 082.C-0333, 082.C-0357, 082.C-0390, 082.C-0412, 082.C-0427, 082.C-0608, 082.C-0718, 083.C-0186, 083.C-0413, 083.C-0794, 083.C-1001, 083.D-0668, 084.C-0185, 084.C-0228, 084.C-0229, 084.C-1039, 085.C-0019, 085.C-0063, 085.C-0318, 085.C-0393, 086.C-0145, 086.C-0230, 086.C-0284, 086.D-0240, 087.C-0012, 087.C-0368, 087.C-0649, 087.C-0831, 087.C-0990, 087.D-0511, 088.C-0011, 088.C-0323, 088.C-0353, 088.C-0513, 088.C-0662, 089.C-0006, 089.C-0050, 089.C-0151, 089.C-0415, 089.C-0497, 089.C-0732, 089.C-0739, 090.C-0395, 090.C-0421, 090.C-0540, 090.C-0849, 091.C-0034, 091.C-0184, 091.C-0271, 091.C-0438, 091.C-0456, 091.C-0471, 091.C-0844, 091.C-0853, 091.C-0866, 091.C-0936, 091.D-0469, 092.C-0282, 092.C-0454, 092.C-0579, 092.C-0721, 092.C-0832, 092.D-0261, 093.C-0062, 093.C-0409, 093.C-0417, 093.C-0474, 093.C-0919, 094.C-0090, 094.C-0297, 094.C-0428, 094.C-0797, 094.C-0894, 094.C-0901, 094.C-0946, 094.D-0056, 094.D-0596, 095.C-0040, 095.C-0105, 095.C-0367, 095.C-0551, 095.C-0718, 095.C-0799, 095.C-0947, 095.D-0026, 095.D-0717, 096.C-0053, 096.C-0082, 096.C-0183, 096.C-0210, 096.C-0331, 096.C-0417, 096.C-0460, 096.C-0499, 096.C-0657, 096.C-0708, 096.C-0762, 096.C-0876, 096.D-0402, 096.D-0717, 097.C-0021, 097.C-0090, 097.C-0390, 097.C-0434, 097.C-0561, 097.C-0571, 097.C-0864, 097.C-0948, 097.C-1025, 097.D-0156, 097.D-0717, 098.C-0269, 098.C-0292, 098.C-0304, 098.C-0366, 098-C-0518, 098.C-0518, 098.C-0739, 098.C-0820, 098.C-0860, 098.D-0717, 099.C-0093, 099.C-0138, 099.C-0205, 099.C-0303, 099.C-0304, 099.C-0374, 099.C-0458, 099.C-0491, 099.C-0798, 099.C-0880, 099.C-0898, 099.D-0717, 1101.C-0721, 180.C-0886, 183.C-0437, 183.C-0972, 183.D-0729, 184.C-0639, 184.C-0815, 185.D-0056, 188.C-0265, 188.C-0779, 190.C-0027, 191.C-0505, 191.C-0873, 192.C-0224, 192.C-0852, 196.C-0042, 196.C-1006, 198.C-0169, 198.C-0836, 198.C-0838, 281.D-5052, 281.D-5053, 282.C-5034, 282.C-5036, 282.D-5006, 283.C-5017, 283.C-5022, 288.C-5010, 292.C-5004, 295.C-5031, 495.L-0963, 60.A-9036, 60.A-9700, and 63.A-9036.
This research has made use of the SIMBAD database, operated at CDS, Strasbourg, France. This research has made use of the NASA Exoplanet Archive, which is operated by the California Institute of Technology, under contract with the National Aeronautics and Space Administration under the Exoplanet Exploration Program.
\end{acknowledgements}

\begin{appendix} 
\section{Estimating the average number of planets per star} \label{AppendixA}
Under the most simplistic assumptions, the number of planets $\Npl$ around each star (within a range of planet/stellar properties), would follow a Poisson distribution, with a rate parameter $\fpl$: $\Npl \sim \mathrm{Pois}(\fpl)$ \citep{Har2001}. We have implicitly assumed that different planets around the same star occur independently. Although this should not be true in general, it is a common assumption used in studies of the planet-occurrence rate  \citep[e.g.][]{Fulton, Hsu18}.

\begin{figure}
\centering
\includegraphics[width=8cm]{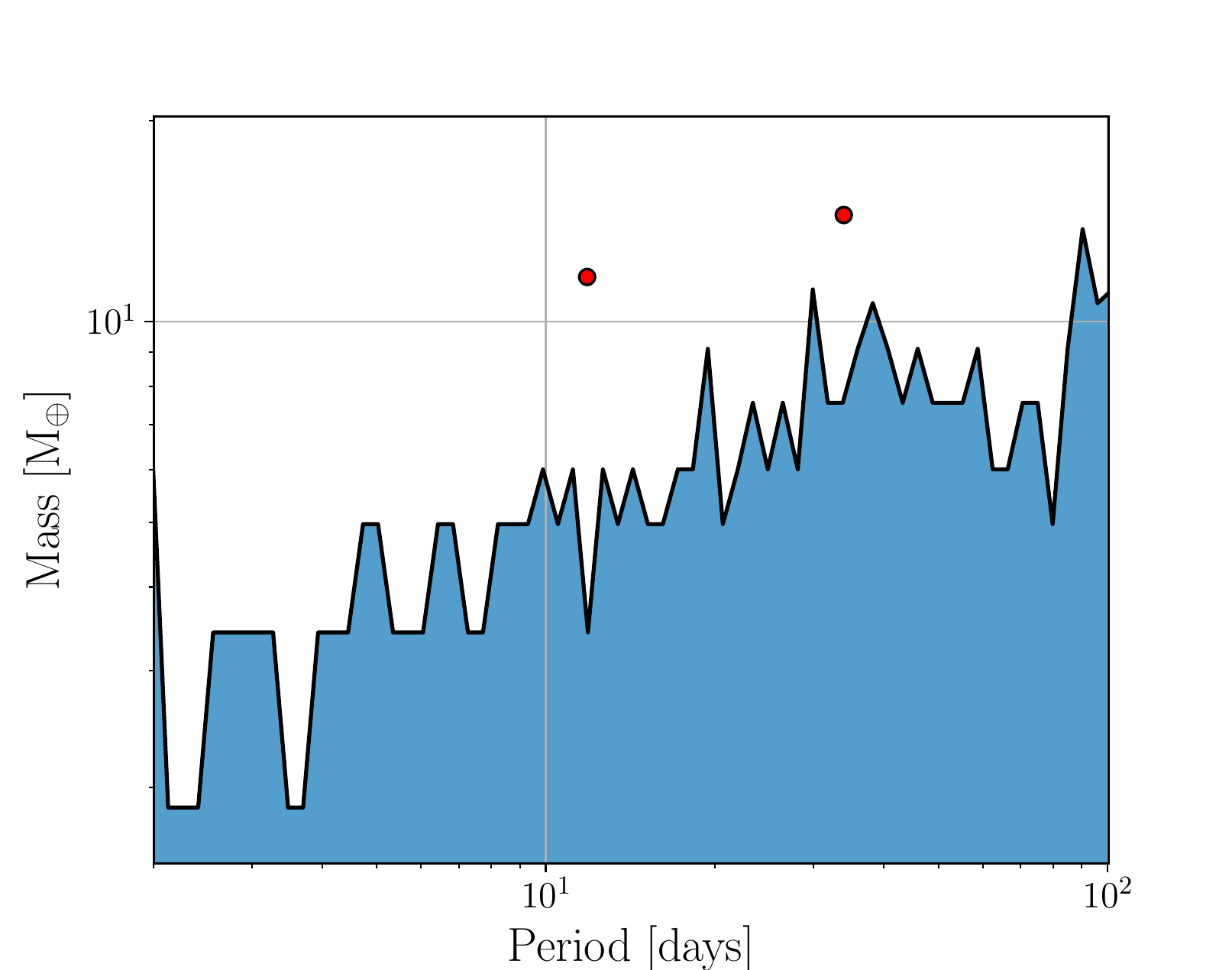}
  \caption{Mass--detection limit curve (calculated according to the LPA method \citep{Meunier} of the planetary system HD~20003 with its two confirmed planets \citep{Udry2019}. We approximate the planet detection probability for this star ($p_{\mathrm{S}} = 0.48$) as the ratio between the area above its mass detection limit curve (white area) and the total area of the rectangle in the period-mass range of $2$--$100$ days and $1$--$20\,M_{\oplus}$.}
     \label{DetectionProbFig}
\end{figure}

For the purpose of Bayesian inference, we can assign a conjugate prior to the rate parameter $\fpl$ in the form of a gamma distribution \citep{RaiSch1961} with shape parameter $\alpha_0$ and rate parameter $\beta_0$. The posterior distribution of $\fpl$ would then also be a gamma distribution with shape parameter $\alpha_0 + \Ntotpl$ and rate parameter $\beta_0 + \Neff$:
   \begin{equation}
      p(\fpl | \Ntotpl, \Neff) \sim \mathrm{Gamma}(\alpha_0 + \Ntotpl, \beta_0 + \Neff) \, ,
   \end{equation}
where $\Ntotpl$ is the number of planets detected around stars in the sample, and $\Neff$ is the effective number of stars searched, which is estimated as the sum of the planet detection probabilities of all the stars: $\Neff = \sum_{\mathrm{i=1}}^{\Ns} p^{\mathrm{i}}_{\mathrm{S}}$ \citep{Hsu18}. Based on the LPA formalism, we define the planet detection probability of a star by the ratio between the area above its mass-detection-limit curve and the total area of the rectangle of logarithmic planetary mass and period (see the example in Fig. \ref{DetectionProbFig}). 

For non-zero detection cases, we report in the main text the posterior median and the $16$\%--$84$\% confidence interval. However, it is quite informative to examine the resulting mean and standard deviation of the posterior gamma distribution:
   \begin{equation}
      \label{eq:munp}
      \mu_{\fpl} = \frac{\alpha_0 + \Ntotpl}{\beta_0 + \Neff} \,,
   \end{equation}
   \begin{equation}
      \label{eq:signp}
      \sigma_{\fpl} = \frac{\sqrt{\alpha_0 + \Ntotpl}}{\beta_0 +\Neff} \,.
   \end{equation}

The minimal prior knowledge we can safely assume regarding $\fpl$ is that it is a positive real value. We wish to avoid imposing any preferred scale on $\fpl$. The natural way to achieve this is by using the log-uniform distribution. In principle, this distribution requires specification of the upper and lower bounds of its support that would still impose some arbitrary scale. Assuming a lower bound of zero and no upper bound gives rise to a log-uniform distribution over the set of positive reals, whose integral does not converge. It is thus an improper prior, which is nevertheless commonly used in Bayesian inference, as long as it is used cautiously, only as a prior distribution, and not over-interpreted as a distribution in its own right \citep[e.g.][]{TarLin2010}. Fortunately, this improper distribution can also be described as a gamma distribution with vanishing shape and rate parameters $\alpha = \beta = 0$, values that are not allowed for proper gamma distributions. Therefore, we substitute in Eq.~\ref{eq:munp} and \ref{eq:signp} $\alpha_0=\beta_0=0$, which leads to quite intuitive final estimates: the posterior mean is simply the ratio of the number of planets detected to the effective number of stars around which planets could have been found.

The formalism above, with the choice $\alpha_0=\beta_0=0$, fails in the case of zero detections. The resulting shape parameter of the posterior distribution would turn out to be zero, which is not a permissible value for the shape parameter, as it leads to a non-integrable posterior probability density function. The case of zero detections poses a problem even for a simple, frequentist, maximum-likelihood estimation, because the resulting Poisson rate parameter would be zero, which is not permissible also for Poisson distribution. Therefore, the case of zero detections calls for a special treatment.

\section{Estimating the fraction of stars with planets} \label{AppendixB}
Under no additional assumptions, the probability of finding $N_{\mathrm{h}}$ planet-hosting stars (within a specified range of planet/stellar properties) among $ \Ns$ stars can be assumed to follow a binomial distribution: $N_{\mathrm{h}} \sim B(\FplH, \Ns)$, where $\FplH$ is the fraction of stars with planets \citep{Har2001}. 
The conjugate prior of a binomial distribution is a beta distribution \citep{RaiSch1961}, parameterized by the two shape parameters $\tilde{\alpha_0}$ and $\tilde{\beta_0}$. The posterior distribution for $\FplH$ will then be a beta distribution with shape parameters $\tilde{\alpha_0} + \NtotplHost$ and $\tilde{\beta_0} + \Neff-\NtotplHost$:
   \begin{equation}
      p(\FplH | \NtotplHost, \Neff) \sim \mathrm{Beta}(\tilde{\alpha_0} + \NtotplHost, \tilde{\beta_0} + \Neff-\NtotplHost) \,,
   \end{equation}
where $\NtotplHost$ is the number of planet-host stars in a sample.

Similarly, the resulting mean and standard deviation of the posterior beta distribution of $\FplH$ are:
   \begin{equation}
      \label{eq:mufh}
      \mu_{\FplH} = \frac{\tilde{\alpha_0} + \NtotplHost}{\tilde{\alpha_0} +\tilde{\beta_0} +\Neff} \,,
   \end{equation}
   
   \begin{equation}
      \label{eq:sigfh}
      \sigma_{\FplH} = \sqrt{\frac{(\tilde{\alpha_0} + \NtotplHost) \cdot (\Neff - \NtotplHost + \tilde{\beta_0})}{(\Neff + \tilde{\alpha_0} + \tilde{\beta_0})^2 \cdot (\Neff + \tilde{\alpha_0} + \tilde{\beta_0} + 1)}}
      \,.
   \end{equation}
The least restrictive prior for $\FplH$ is obviously a uniform distribution between $0$ and $1$, which is equivalent to a beta distribution with $\tilde{\alpha} = \tilde{\beta} = 1$. As is evident from Eq.\ \ref{eq:mufh} and \ref{eq:sigfh}, for sufficiently large samples, where $\NtotplHost,\Neff \gg 1$, the posterior mean of the fraction of stars with planets can be conveniently approximated as the ratio of the number of planet-host stars to the effective stellar sample size. Again, in the main text we report the median and the 14\%--86\% confidence interval.

\end{appendix}

\end{document}